\documentclass[aip,reprint,graphicx]{revtex4-1}
\pdfoutput=1
\usepackage{mathtools}
\usepackage{graphicx}
\usepackage{gensymb}
\usepackage[export]{adjustbox}
\usepackage{amsmath}
\usepackage{float}
\usepackage{hyperref}
\usepackage{hypcap}
\draft

\begin{document}

\title{Effects of hydrophobic solute on water normal modes}

\author{Kambham Devendra Reddy}
\author{Albin Joy}
\author{Rajib Biswas}
\email[Electronic mail: ]{rajib@iittp.ac.in}
\affiliation{Department of Chemistry and Center for Atomic, Molecular and Optical Sciences \& Technologies, Indian Institute of Technology Tirupati, Yerpedu 517619, Andhra Pradesh, India}

\begin{abstract}
The vibrational modes of water get significantly modified by external solutes; this becomes particularly important when the solute is hydrophobic. In this work, we examine the effects of a tiny hydrophobe, methane, on the normal modes of water, using small cluster-based harmonic normal mode analysis of aqueous methane system. We estimate the vibrational density of states and also the infrared spectral density. We compare the methane-water data with the bulk water response. We decompose these modes based on different vibrational characters. The stretch-bend decomposition reflects a pronounced coupling between the methane asymmetric stretch mode and the water symmetric stretch mode. We examine the methane-water data in terms of the symmetry of the central water molecule's vibrations and find that asymmetric modes do not contribute. We also find that the vibrational modes having non-zero contributions from methane molecule are extremely localized in nature.
\end{abstract}
\maketitle 

\section{Introduction}
Hydrophobic hydration plays a fundamental role in several biological processes such as membrane formation, protein folding, ligand binding, assembly of proteins into functional 
complexes, lipid bi-layer formation, and many others \cite{Tanford1973,Srinivas2002,Srinivas2003,Mason2004,Chandler2005,Dyson2006,Bakulin2011,Hazra2014,Ben2015,Grdadolnik2017}. It 
is also evident that the abundant hydrogen bond (H-bond) network of bulk water gets significantly modified in the presence of different solutes, particularly important in the case 
of hydrophobic molecules. The influence of such molecules on the structure of liquid water is an essential topic of ongoing research. The quest for hydrophobic hydration has emerged 
with the seminal work of Frank and Evans \cite{Frank1945} who anticipated that “\textit{water forms frozen patches or microscopic icebergs around such solute molecules.}” The low solubility of 
any such oily solutes poses significant experimental challenges to probe the system experimentally. Thus a numerous earlier explorations of hydrophobic hydration in the case of the 
classic hydrophobe, methane ($\mathrm{CH_4}$), have been done mainly by employing theoretical,\cite{Kauzmann1959} molecular dynamics (MD) \cite{Rossky1982,Laza1992,Ash1996,Sharp1997,Hummer1998,Pas2009},
and ab initio (AIMD) simulation techniques \cite{Mon2012,Galamba2014,Grdadolnik2017}.
\par
A few experimental findings of hydrophobic hydration portray conflicting conclusions
\cite{Dejong1997,Buchanan2005,Laage2009,Perera2009,Tie2010,Str2014,Rankin2015,Grdadolnik2017,Ben2018}. The neutron diffraction study by De Jonget al. concluded that 19 tangentially oriented water molecules surround the methane molecule \cite{Dejong1997}, and a similar experiment by Buchananet al. concluded “\textit{no evidence for enhanced water structure.}” \cite{Buchanan2005} A recent work by Grdadolnik $\textit{et al}$. using infrared (IR) spectroscopy of isotopically dilute systems in addition to ab initio MD suggested 
“\textit{evidence that supports, unequivocally, the iceberg view of hydrophobicity.}” This is inconsistent with other classical MD \cite{Rossky1982,Laza1992,Ash1996,Sharp1997,Hummer1998,Pas2009} and 
AIMD \cite{Mon2012,Galamba2014,Grdadolnik2017} data which find no icebergs around hydrophobic molecules. However, the MD hydration thermodynamic analyses \cite{Qvist2008} exactly reproduce the entropic anomalies that originally encouraged the Frank and Evans iceberg hypothesis \cite{Frank1945}. Furthermore, experimental fs-IR \cite{Rezus2007,Petersen2009,Str2014,Rankin2015,Ben2018}, NMR \cite{Laage2009}, and theoretical explorations \cite{Ros2012,Mon2012} find slower water dynamics around hydrophobic moieties, with no evidence of iceberg formation. Recent Raman multivariate curve resolution (Raman-MCR) 
hydration-shell vibrational spectroscopic measurements find that “\textit{methane’s hydration-shell is slightly more clathrate-like than pure water at ambient, and lower temperatures.}” \cite{Ben2018,Ben2019,Bredt2020}
\par
Despite the previous conflicting conclusions, it is evident that the underlying many-body interactions could give rise to different hydration arrangements around the hydrophobe \cite{Wiggins1997,Lum1999,Huang2000,Chandler2005,Bakulin2011,Galamba2013,Galamba2014,Grdadolnik2017}. This modifies the strength of the hydrogen bonds in the close vicinity of the hydrophobe \cite{Scatena2001,Raschke2005,Buchanan2005,Bakulin2011,Davis2012,Mon2012,Grdadolnik2017}. In one of the earlier works, Stillinger has argued that for sufficiently weak solute-water attraction, a large smooth hydrophobic surface might be enclosed by a microscopically thin film of water vapor \cite{Stillinger1973}. 
\par
The high sensitivity of the IR spectroscopy to local structures makes it an appealing method for exploring of these systems, where the hydrophobic moieties alter the water H-bond significantly \cite{Fecko2003,Devendra2020}.
IR spectroscopy of O–H stretching mode serves as the most reliable and sensitive probe for investigating the relative strengths of hydrogen bonds \cite{Davis2012,Fecko2003,Hecht1992,Hecht1993,Sharp1997,Auer2007,Laage2009,Stiopkin2011,Biswas2013,Biswas2016}. Despite its extensive use, the superposition of several solvation 
structures embedded in the experimental data makes it extremely difficult to establish the structure-spectrum correlation accurately. Besides, a further delocalization arises due to the strong inter molecular an-harmonic couplings. In contrast, the molecular level resolution of computer simulation aided 
theoretical spectroscopic techniques enables us to probe these systems at the microscopic level \cite{Auer2007,Biswas2013,Biswas2016,Corcelli2004,Corcelli2005,Roberts2009,Biswas2017,Samanta2018}. 
In several earlier works, mixed quantum-classical (MQC) spectroscopy models have been extensively used to study the isotope dilute aqueous systems using O–H stretching vibrations as probe \cite{Auer2007,Biswas2013,Biswas2016,Corcelli2004,Corcelli2005,Samanta2018}. However, this reduced dimensional description does not allow us to investigate the extent of mode mixing 
in the condensed phase. The instantaneous normal mode analysis is found to be extremely useful in this regard \cite{Biswas2017,Cho1994}. 
In this work, we explore the methane-water system's vibrational responses using instantaneous normal mode analyses of small methane-water clusters drawn from classical MD trajectories.
A particular emphasis is given to understand the 
effects of the hydrophobic molecule on the mode-mixing and the extent of delocalization.
\par
The rest of this paper is organized as follows: we discuss the simulation details, cluster selection, and normal mode calculation details in Section II, Section III includes the results 
and discussions, and conclusions are given in Section IV.
\section{Simulation Details}
We use the Groningen machine for chemical simulations (GROMACS) version 2019.1 \cite{Abraham2015} to perform classical molecular dynamics simulations of two systems: (1) methane-water system 
consisting of one methane molecule and 255 water molecules, and (2) bulk water system with 256 water molecules. We use optimized potentials for liquid simulations all-atom (OPLS-AA) \cite{Jorgensen1996} 
force field to describe the methane molecule, and extended simple point charge (SPC/E) \cite{Berendsen1987} force field to describe the water molecules. We employ periodic boundary conditions in all 
three directions. First, we perform energy minimization of the systems using the steepest descent algorithm, followed by 1 ns equilibration in the \textit{NPT} ensemble. Thereafter, the 5 ns production 
run is carried out in the \textit{NVT} ensemble. We use Berendsen thermostat \cite{Berendsen1984} with a relaxation time of 0.1 ps, and Parrinello–Rahman barostat \cite{Parrinello1981,Nose1983} with a relaxation time of 
1.0 ps for maintaining a constant temperature at 300 K, and pressure at 1 bar, respectively. For neighbour searching, and non-bonded interactions, we use a 9 {\AA} cut-off radius, and all the bonds are kept 
fixed using linear constraint solver (LINCS) \cite{Hess1997} algorithm. We use Particle Mesh Ewald (PME) \cite{Darden1993} with an FFT grid spacing of 1.6 {\AA} for the long-range electrostatic interaction’s 
calculation.\par
\begin{figure}[htb!]
\centering
\includegraphics[width=0.45\textwidth]{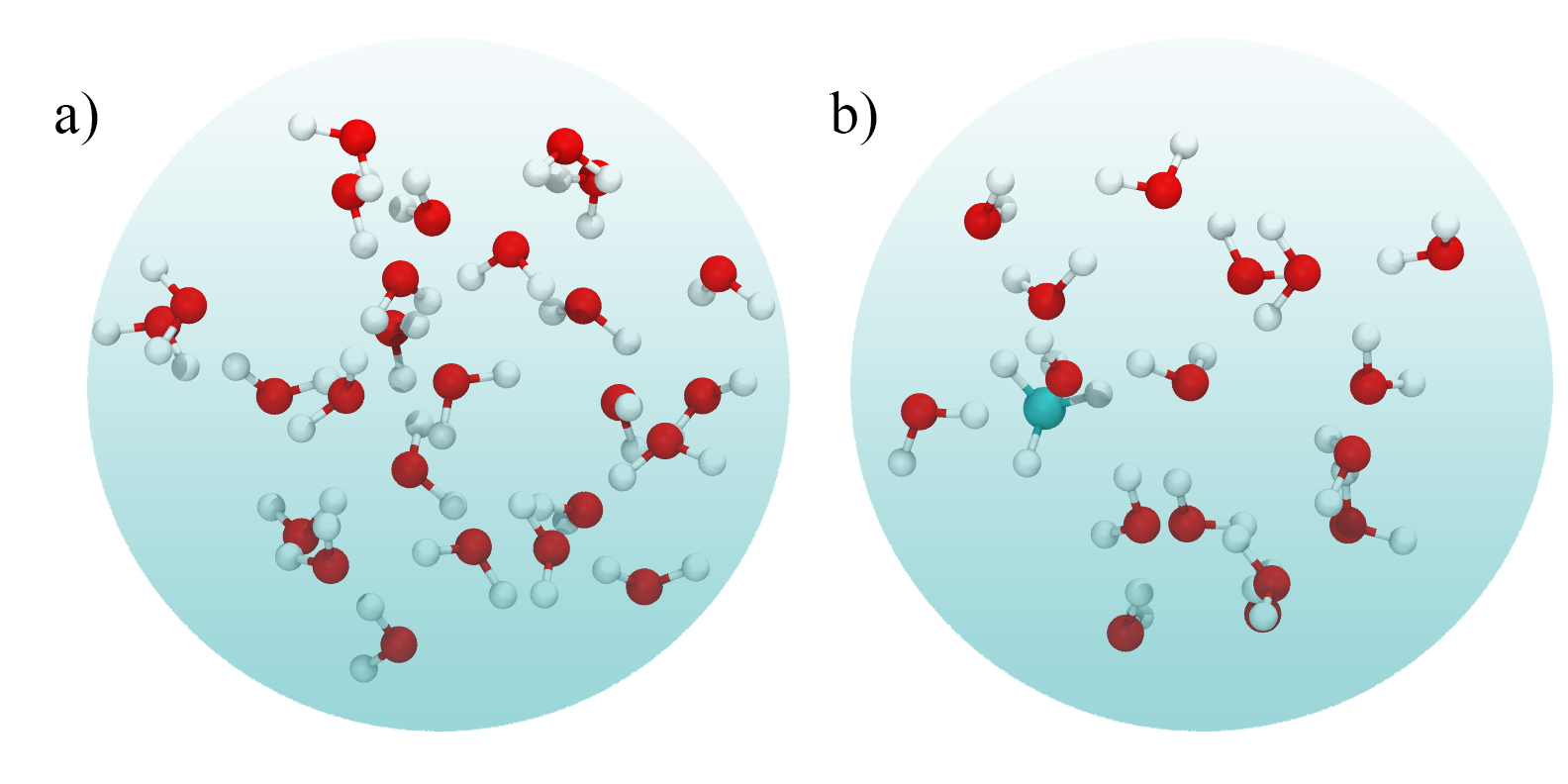}
\caption{\label{fig1} Illustration of clusters of (a) bulk water, and (b) methane-water having 5 {\AA} radius cut-off from the central water molecule.}
\end{figure}
We perform the harmonic normal mode analysis for both systems by employing the following strategy. We utilize electronic structure calculations on instantaneous small frozen clusters extracted from 
the classical MD trajectory. From the classical trajectory, we extract small clusters by identifying a central O atom that belongs to the water molecule closest to the methane molecule and including all 
water molecules having their oxygen atoms within a 5.0 {\AA} radius (Figure \ref{fig1}). We obtain the methane-water data by using such 2000 methane-water clusters. We choose 1500 small clusters extracted similarly from the bulk water trajectory to extract the bulk water response. The methane-water clusters contain an average of ~16 water molecules, and bulk water clusters have an average 
of ~18 water molecules, which are adequate to approximately resemble the bulk-like environment around the central water molecule. Minor variations in the number of water molecules in a cluster do not 
alter the conclusions.
\par
We sample methane-water clusters based on the distance $(r_\mathrm{MW})$ between the methane and the nearest water molecule i.e. the central water molecule. To ensure that these clusters sample a wide variety of relevant configurations, we select the clusters with $r_\mathrm{MW}$ within 2.80 {\AA} – 4.00 {\AA}. For bulk water clusters, we tune the distance between the central water molecule and its nearest water molecule $(r_\mathrm{WW})$ and the selected clusters have $r_\mathrm{WW}$ within 2.40 {\AA} – 3.10 {\AA}. We perform harmonic normal mode analysis based on density functional theory (DFT) calculations by using 
B3LYP \cite{Vosko1980,Lee1988,Becke1993,Stephens2002} hybrid functional in the Gaussian 16 software \cite{Frisch2016}, and 6-311++G(d,p) basis set without any geometrical optimization or any other constraints.
We calculate the normal modes by constructing the Hessian matrix from the potential energy obtained by single-point energy calculations. From the eigenvalues of this Hessian matrix, we obtain the harmonic frequencies. 
The displacements of the atoms involved in the i-th normal mode are obtained from the eigenvectors, ${e_{i}}$. As the instantaneous snapshots from the classical MD are not geometrically optimized, we obtain ~7 {\%} 
low-frequency range imaginary modes and are excluded from our analysis. Finally, we separate all possible real normal modes and corresponding intensities. The intensity of any normal mode is calculated using the 
following expression 
\begin{equation}
\label{eq:intensity}
  I(\omega_{i}) =  {\left|\frac{d{\mu}}{dq_{i}}\right|}^2 
\end{equation}where ${\mu}$ represents dipole moment and ${q_{i}}$ is the displacement of the \textit{i}-th normal mode. We evaluate the density of states (DOS), ${D(\omega)}$ and intensity-weighted density of states (IDOS) i.e. spectral density, ${\rho(\omega)}$, by using the following equations. 
\begin{equation}
\label{eq:dos}
   \ D(\omega) =  \frac{1}{N_\mathrm{{clusters}}}\sum_{\mathrm{clusters}} \: \sum_{\mathrm{modes} \: i} \delta({\omega}-{\omega_i})
\end{equation}\begin{equation}
\label{eq:idos}
   \rho(\omega) =  \frac{1}{N_{\mathrm{clusters}}}\sum_{\mathrm{clusters}} \: \sum_{\mathrm{modes} \: i} I({\omega_i})\delta({\omega}-{\omega_i})
\end{equation}We obtain the density of states, $D({\omega})$, by constructing a normalized histogram of vibrational frequencies of all normal modes binned to windows of 20 cm{$^{-1}$} and plot as a function of the bin center frequencies. We construct the spectral density from the same histogram by weighting each normal mode with the corresponding intensity.

\section{Results and Discussion}
\subsection{Full dimensional response}
\begin{figure}[htb!]
\centering
\includegraphics[width=\linewidth]{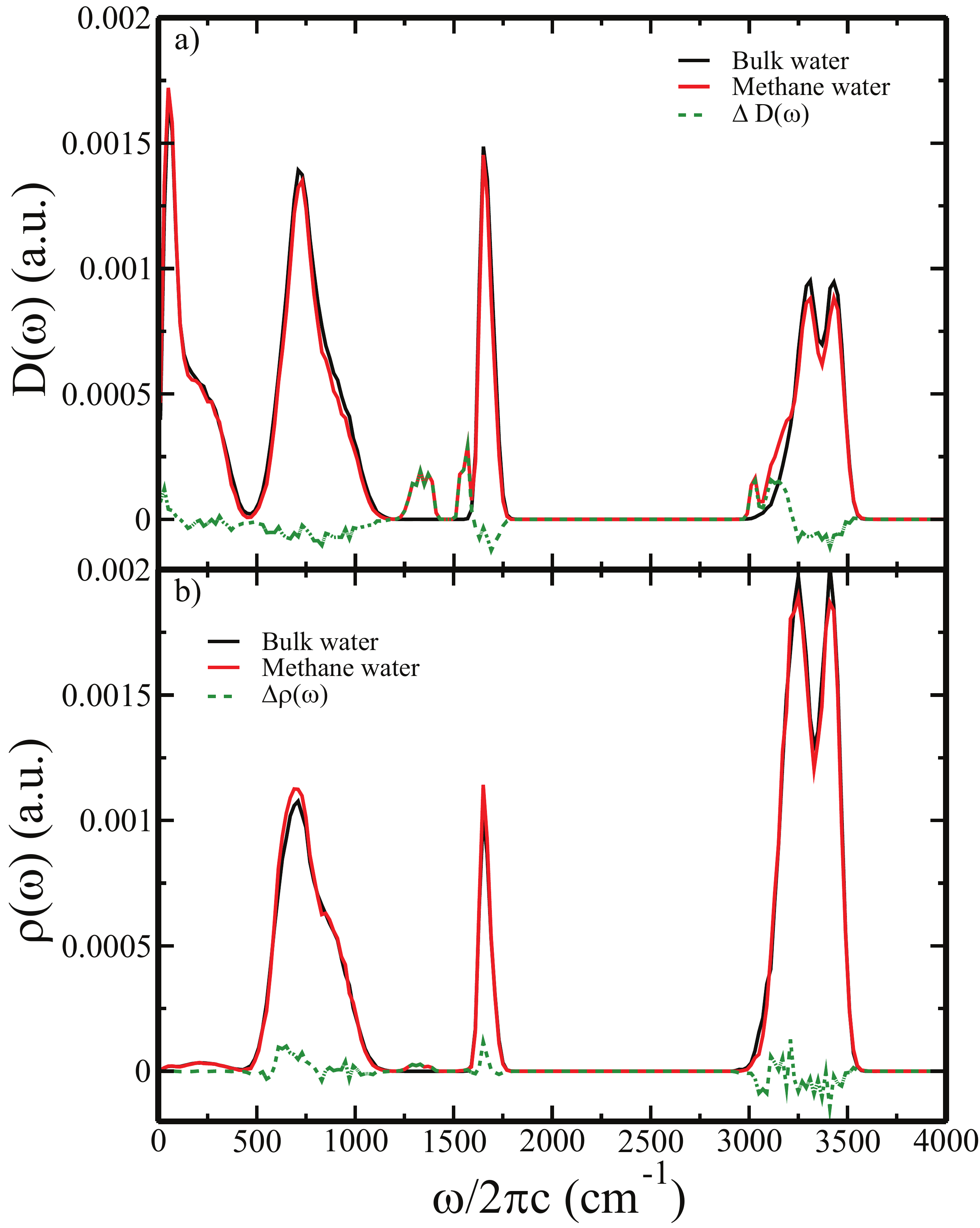}
\caption{\label{fig2} (a) The density of states, ${D(\omega)}$ of bulk water (black), methane-water (red), and the difference in density of states, ${\Delta D(\omega)}$ of methane-water relative to bulk water (green). (b) The spectral density, ${\rho(\omega)}$ of bulk water (black), methane-water (red), and the difference in spectral density of methane-water relative to bulk water (green).}
\end{figure}
The instantaneous normal mode analysis provides microscopic information about the mixing of different modes, the relative symmetry of normal modes, and the extent of delocalization of normal mode vibrations \cite{Cho1994}. We represent the DOS, ${D(\omega)}$ and the IDOS, ${\rho(\omega)}$ of bulk water and methane-water clusters in Figure \ref{fig2}. These data include all modes, even modes from cluster surface vibrations. It is clear from Figure \ref{fig2}, that the DOS of bulk water reproduced almost all the features as reported earlier \cite{Biswas2017}. The normal mode analysis of only water clusters predicts the shifts in peak position and line-width relative to the experiment \cite{Biswas2017}. Despite these limited dissimilarities, one can easily recognize the qualitative similarities with the experimental spectrum. The center of the stretch region is arising at a lower frequency (3340 vs. 3400 cm$^{-1}$) and wide as compared to the experimental spectrum (314 vs. 265 cm$^{-1}$). The peak at 1650 cm$^{-1}$ is the closest to the experimental data and is found to be subjugated by the water bending character. The librational motions are observed at a significantly higher frequency, 715 cm$^{-1}$, than that observed in the experiment. These observations corroborate that DFT calculations considerably overestimate the strength of the hydrogen bonding interaction in water clusters \cite{Biswas2017}. Another difference we observe in the present case is that a bimodal distribution arises in the case of water stretching modes. This might originate from the slight difference in nuclear configurations sampled by the rigid water model as compared to the flexible water model \cite{Biswas2017}. The symmetric stretching modes show a peak maximum at 3300 cm$^{-1}$, and the anti-symmetric stretching modes show a peak maximum at 3430 cm$^{-1}$.  In the methane-water system, we find three new signatures in the DOS. One peak arises at 1350 cm$^{-1}$ because of the umbrella motion of 3H atoms about the other C-H bond. The peak around 
1560 cm$^{-1}$ is originating from the bend motion of the methane molecule, and another high-frequency signature arises in the frequency range of 3000–3200 cm$^{-1}$, from the anti-symmetric C-H stretching mode. Besides, for the methane-water system, a decrease in intensity is also observed throughout the water stretch region compared to the bulk water data. The lower intensity is reflecting the reduction in the number of average water molecules in the case of methane-water clusters. The IDOS, $\rho(\omega)$  of bulk water and methane-water systems are shown in Figure \ref{fig2}b. 
As reported earlier \cite{Biswas2017}, we observe a similar trend in bulk water response except for the bimodal distribution in the stretch region. In the case of methane-water response, we indeed observe 
different features as predicted by the DOS data. However, the respective methane-active normal mode's low intensity gives rise to a less prominent signature in the spectral density.
\par
\begin{figure}[htb!]
\centering
\includegraphics[width=\linewidth]{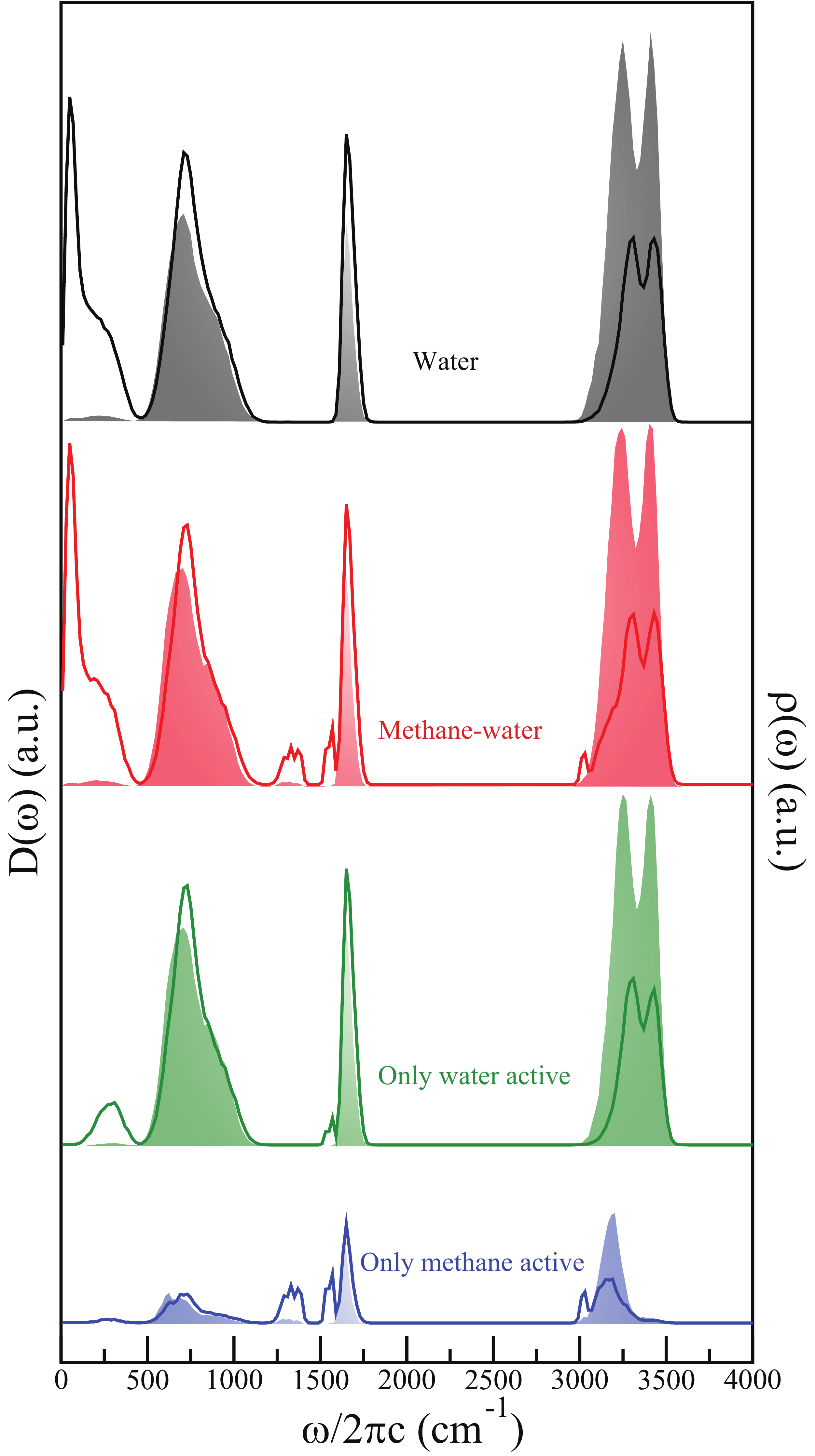}
\caption{\label{fig3} Density of states (solid lines), and spectral density (gradient fill) for bulk water (black), methane-water (red), only water active (green), only methane active (blue) modes. Only water active modes are isolated based on displacement of the atoms in the central water molecule $\ge 0.01$ {\AA}, and displacement of atoms in methane molecule $< 0.01$ {\AA}. The only methane active modes are obtained by using the reversed condition.}
\end{figure}
Furthermore, we investigate the difference in DOS, $\Delta D(\omega)$, and that in IDOS, $\Delta \rho(\omega)$ to probe the effect of the non-polar methane moiety on the distributions of DOS and IDOS more clearly. 
We obtain $\Delta D(\omega)$ and $\Delta \rho(\omega)$ by subtracting the water response from the methane-water response. In both cases, we observe negative values which signify the loss of a few water molecules in the methane-water clusters. We observe clear signatures for methane active normal modes in the case of $\Delta \rho(\omega)$.
\par
In order to isolate the vibrations of the central water molecule from the contributions of surface water, we have selectively calculated the DOS and IDOS for those normal modes in which only the central water molecule has non-zero contribution. To achieve this, we track the normal mode displacement vector and isolate those normal modes in which the central water molecule has a non-zero 
displacement (typically $\ge$ 0.01 {\AA}) (see supplementary material).
\par
We also isolate normal modes based on the participation of the methane molecule and the central water molecule. We track the normal mode displacement of these two molecules to isolate the modes in which only the methane molecule is active and only the central water molecule is active. Only water active modes are isolated based on the displacement of the atoms in the central water molecule $\ge$ 0.01 {\AA} and the displacement of the atoms in the methane molecule $<$ 0.01 {\AA}. The only methane active modes are obtained by using the reversed condition. In Figure \ref{fig3}, 
we represent the isolated DOS and the spectral density of methane-water and bulk water systems with various conditions.
\par
\begin{figure*}
\centering
\includegraphics[width=\linewidth]{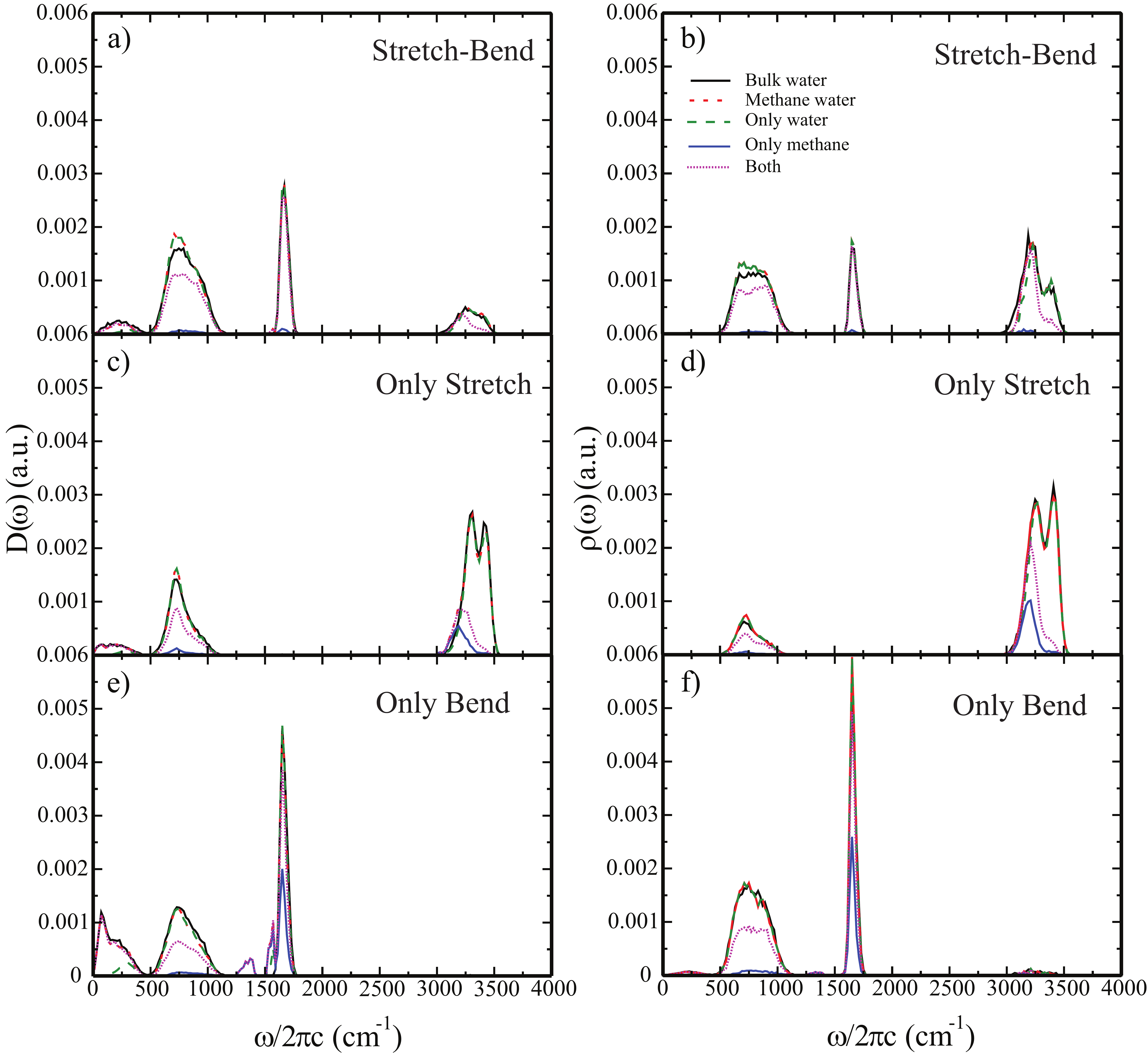}
\caption{\label{fig4} Density of states and spectral density of states of stretch-bend, only stretch and only bend active modes within bulk water, methane-water, only water active, only methane active and both active clusters.  Only stretch peak calculated based on displacement of central water molecule OH bond $\ge 0.01$ {\AA} and only bend peak calculated based on water molecule HOH angle changes $\ge 1^\circ$ }
\end{figure*}
We find the corresponding DOS and IDOS of the bulk water system, and the only water active normal modes in the methane-water system are almost similar (see supplementary material). We observe three additional features in the DOS of the methane-water system, as explained earlier. By separating the only methane active modes, we notice the disappearance of the bimodal stretching frequency, which is present in bulk water, and the only water active data of methane-water clusters. Hence, it is evident that the bimodal distribution of DOS and IDOS in the high-frequency range (3100 - 3550 cm$^{-1}$) originates from the water active modes. Despite an overall reduction of water active modes in the only methane active modes, it is clear from Figure \ref{fig3} that there exists a small coupling of these methane active modes with the motion of other water molecules present in the cluster. To quantify the contribution of water motions in the only methane active data, we have estimated the DOS and IDOS with the displacement of the atoms in the 
methane molecule $\ge 0.01$ {\AA}, and displacement of atoms of all other water molecules $< 0.01$ {\AA}.
\begin{figure}
\centering
\includegraphics[width=\linewidth]{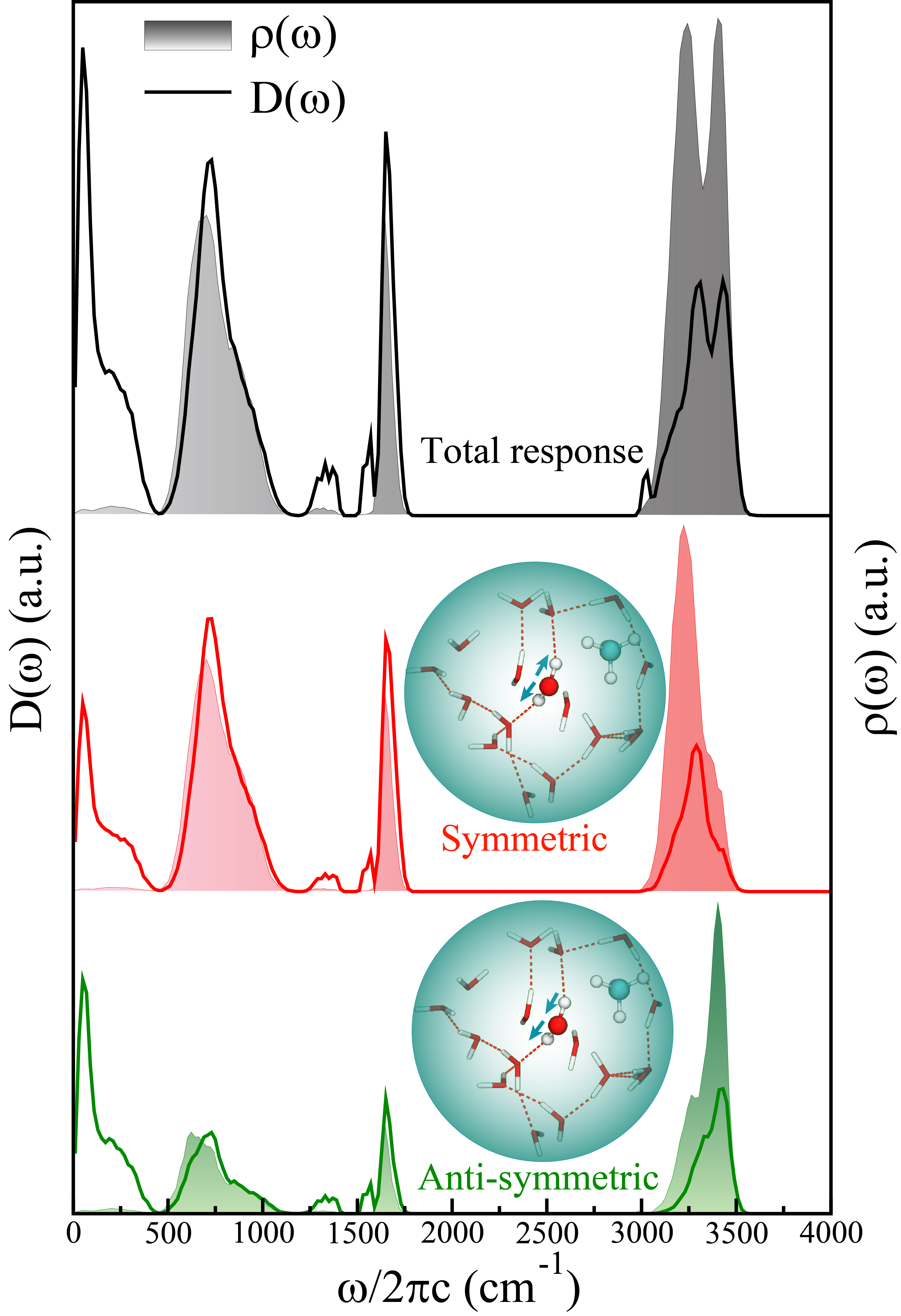}
\caption{\label{fig5} DOS (${D(\omega)}$, solid line) and IDOS (${\rho(\omega)}$, gradient fill) of the methane-water system (black), symmetric (red) and anti-symmetric (green) modes. Symmetric and anti-symmetric modes are separated based on the relative phase of the displacement (shown in the figure) of bonds of the central water molecule.}  
\end{figure}
\subsection{Stretch and bend modes}
To investigate the mixing of stretch and bend vibrations, we decompose the spectral density and DOS by tracking the atomic stretch and bend displacements of the central water molecule for both systems. We consider a normal mode to have a bend character if the change in the H–O–H angle of the central water is $\ge 1^\circ$. We consider a normal mode to have a stretch character if any O–H bond of the central water molecule has a displacement $>$ 0.01 {\AA}. By comparing these two displacements, we subdivide the normal modes into a stretch only, bend only or both stretch and bend characters.\par 
We indeed check for different cut-off values for these two displacements. Of course, the choice of the cut-offs influences the conclusions regarding the extent of mode mixing. However, several qualitative deductions remain unaffected. We select these cut-offs in such a way so that we can observe the mode mixing with higher resolution. The dependence of the degree of mode mixing on the choice of different cut-off values is represented in the supplementary material.\par
The decomposition of modes with only stretch and only bend characters are represented in Figure \ref{fig4}. This data is further decomposed based on the relative participation of the methane and the central 
water molecule. As explained earlier, we decompose only water active modes based on the displacement of the atoms in the central water molecule $\ge$ 0.01  {\AA} and the displacement of atoms in the methane molecule $<$ 0.01 {\AA}. The only methane active modes are obtained by using the reversed condition. In the case of both active modes, we use displacement of the atoms in the central water molecule and the methane molecule to be $\ge$ 0.01 {\AA}.\par

\begin{figure}[htb!]
\centering
\includegraphics[width=\linewidth]{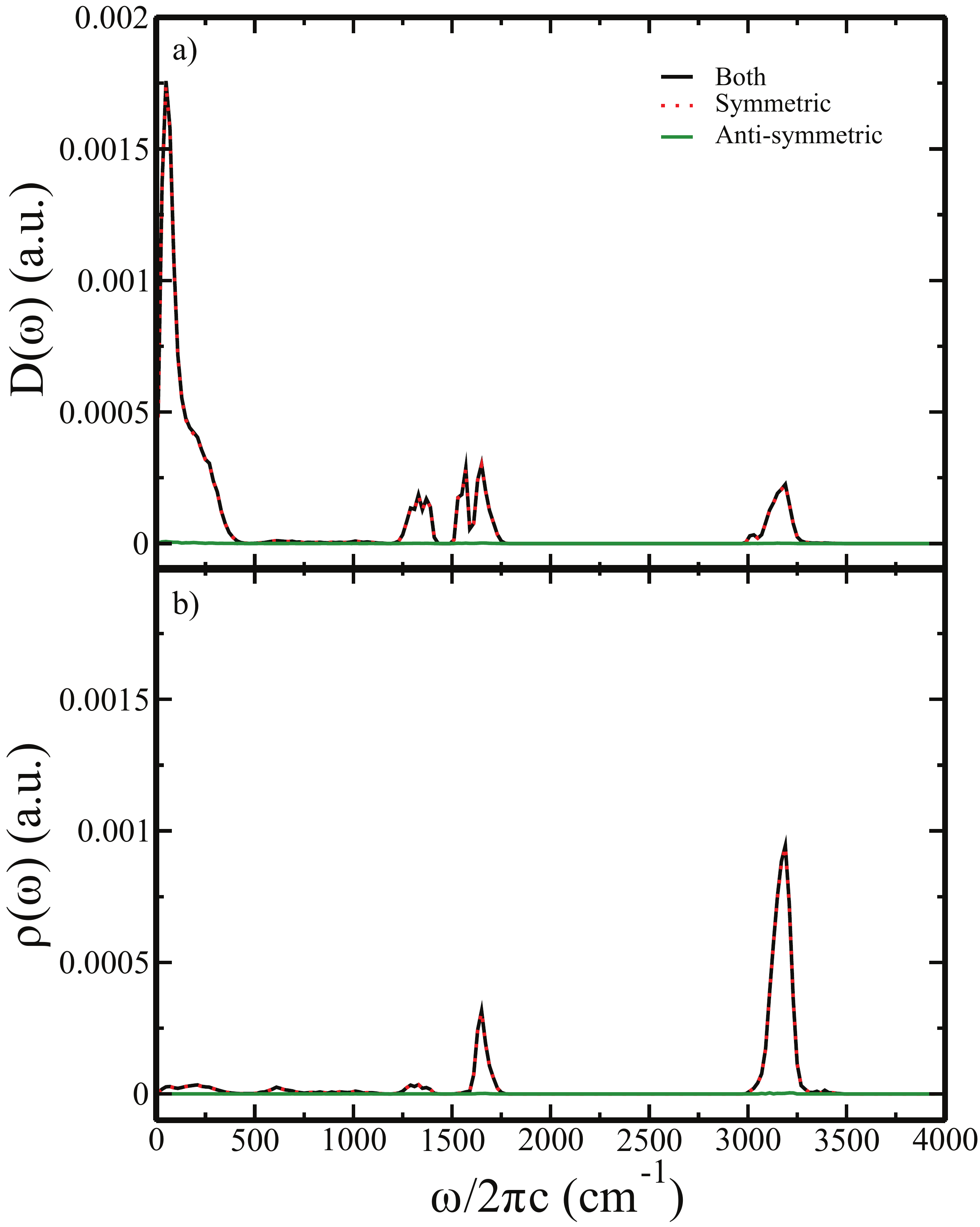}
\caption{\label{fig6} The density of states, D($\omega$) and the spectral density of states, $\rho(\omega)$ for both active (black), symmetric (red) and anti-symmetric (green) modes of methane-water clusters. As described in the main text, both active modes are those which have displacement of the atoms in the central water molecule and the methane molecule $\ge$ 0.01 {\AA}. We further decompose the D($\omega$) and $\rho(\omega)$ based on symmetry of vibrations of the central water molecule in similar fashion as described in Figure \ref{fig5}.}
\end{figure}

It is apparent that the density of only stretch active states in both methane and water active conditions significantly changes. However, the density of only bend active states is not influenced by the presence of methane to a great extent. Interestingly, in the case of both active modes of the stretch-only distribution, the asymmetric stretch contribution is significantly suppressed, and the bimodal feature in the high-frequency range is no longer extant. This indicates that the methane active modes are coupled with symmetric stretch modes of water. To identify the nature of the methane modes, we further perform the symmetry analysis of the different normal modes, and this is presented in section C. Although the `bend only' spectral density of only water active and both methane and water active modes suggests that water bend mode is not getting affected by the methane modes significantly, however, the only methane active modes have small contribution from the other water molecules as well. This suggests that the water bend mode is also coupled with the methane active modes. We also examine the extent of coupling between stretch and bend character by tracking both active modes. This is shown in Figure \ref{fig4}. It is evident from the figure that the stretch and bend character are indeed coupled.
\subsection{Vibrational modes and symmetry}
In this section, we analyze the DOS and IDOS of all the clusters for both systems with respect to the symmetry of vibration of the central water molecule. We track down the relative phase of the motion of the two \mbox{O-H} bonds of the central water molecule. If these \mbox{O-H} bonds are vibrating in phase, we call that mode symmetric mode and anti-symmetric mode otherwise. The definition of symmetric and anti-symmetric modes is shown pictorially in Figure \ref{fig5}. We present the results in Figure \ref{fig5}. The peak at 3330 cm$^{-1}$ corresponds to the symmetric \mbox{O-H} stretching, and that at 3430 cm$^{-1}$ corresponds to the asymmetric \mbox{O-H} stretching (see supplementary information for the bulk water data). It is also clear from the figure that both the symmetric and the asymmetric stretching modes contribute to the water bend and libration modes. However, the contribution from the asymmetric modes is a bit less than that of the symmetric modes. It is also important to realize that in both of these decomposed distributions of DOS and IDOS, we observe the signatures of the methane vibrational motions. Both of these modes also have contributions from the bend and umbrella modes of methane molecule almost equally. However, the asymmetric stretching vibration of methane is more pronounced in the symmetric stretching ensemble of the central water. To further quantify this, we investigate both active mode of methane-water data in terms of the symmetry of vibrations of the central water molecule and found that asymmetric modes do not contribute to the DOS and IDOS shown in Figure \ref{fig6}(see supplementary information). Only methane active data (shown in Figure \ref{fig3}) also suggest the same.
\begin{figure}
\centering
\includegraphics[width=\linewidth]{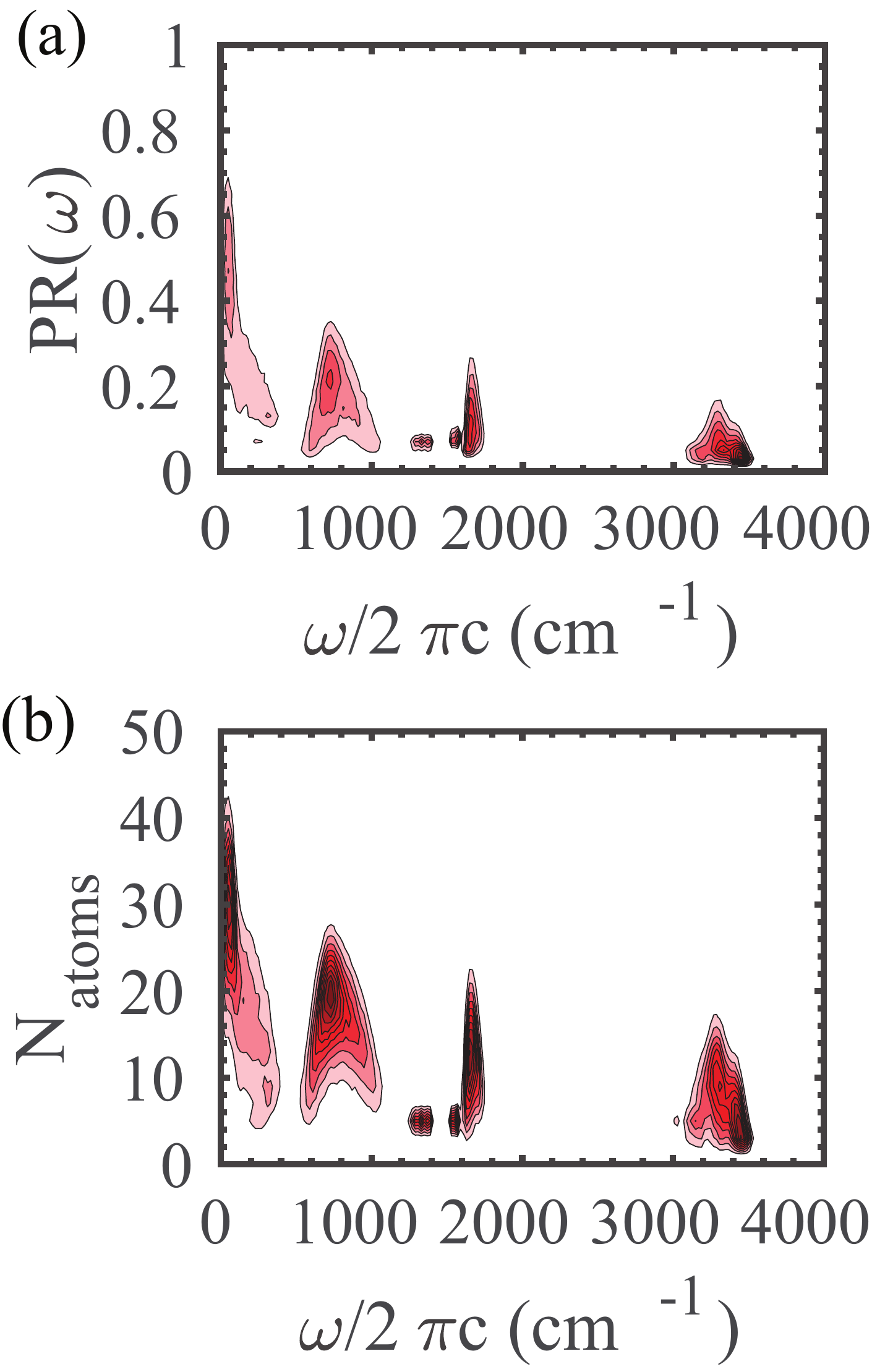}
\caption{\label{fig7} (a) Probability density of participation ratio (PR) of different vibrational modes, and (b) the degree of delocalization in units of number of atoms having any normal mode with ${|e|_{cutoff}=0.1}$ {\AA}.}
\end{figure}
\subsection{Delocalization of normal modes}
To analyze the degree of delocalization of different normal modes, we calculate participation ratio (PR)\cite{Sastry2001} using the following equation.
\begin{equation}
\label{eq:pr}
     \ PR = {{\Big[N\sum_{\alpha=1}^{3N}({e_i^\alpha}{e_i^\alpha})^{2}} \Big]}^{-1}
\end{equation}

where N is the total number of atoms in the cluster, and $\alpha$ refers to the 3N Cartesian coordinates describing the atomic displacements for each normal mode. Thus, for a local mode, the PR is of the order of $N^{-1}$, and approaches unity if it is fully delocalized over the cluster. We also calculate the degree of delocalization by estimating the number of atoms involved in each mode. The following equation calculates the number of atoms involved in the $i^{th}$ mode.
\begin{equation}
\label{eq:dl}
     \ N_\mathrm{atoms}^i = \sum_{j=1}^{N}H({\lvert e_j \rvert}-{\lvert e_{cutoff} \rvert})
\end{equation}

Here ${e_{j}}$ is the displacement vector of the \textit{j}-th atom in a given normal mode, $e_{cutoff}$ is the distance cut-off, and H is the Heaviside step function. We calculated the participation ratio and the degree of delocalization, in units of the number of atoms, for those modes which have vibrational amplitude more than 0.01 {\AA} using the equations \ref{eq:pr} and \ref{eq:dl}. 
Figure \ref{fig7} represents the PR and the degree of delocalization of different modes present in the methane-water clusters. The general observation from the figure is that the low-frequency modes are found to be more delocalized, and the high-frequency modes are localized \cite{Biswas2017}. The low frequency delocalized characters originate from the collective vibrations of the oxygen atoms in the hydrogen-bonded network. In addition to the different water vibrational modes, the figure also depicts the nature of the normal modes in which methane atoms participate actively. It is evident from the figure that the methane umbrella mode, bend mode, and asymmetric stretch modes are highly localized in nature.

\section{Conclusions}
It has been shown earlier that despite certain shortcomings, the cluster-based instantaneous harmonic normal mode analysis predicts different features of the experimental observations qualitatively \cite{Biswas2017}. This method is beneficial for understanding mode-mixing and interpreting the solution-phase IR spectra. In this work, we perform a detailed analysis of the methane-water system and compare the observations with bulk water response. In general, the methane-water system shows a decrease in intensity throughout the water stretch region compared to the bulk water data. This lower intensity reflects the reduction in the number of average water molecules in the case of methane-water clusters. The stretch-bend decomposition of the full DOS and IDOS reflects that only stretch active states in the presence of both methane and water active condition significantly changes, but the presence of methane does not influence the density of only bend active states to a great extent. This indicates a pronounced coupling of the methane asymmetric stretch mode with the water symmetric stretch mode. Also, we decompose the DOS and IDOS based on the symmetry of the vibrations and find that the methane active modes are coupled with symmetric stretch modes of water. We examine both active modes of methane-water data regarding the symmetry of vibrations of the central water molecule and find that asymmetric modes do not contribute to the DOS and IDOS. We investigate the extent of delocalization of different vibrational modes by estimating the participation ratio and also by identifying the number of atoms involved in any mode. This reveals that the vibrational modes having non-zero contributions from methane molecule are extremely localized in nature. 
\par
\vspace{12pt}
\hspace{-18pt}\textbf{Declaration of Competing Interest}\par
\vspace{12pt}
The authors declare that they have no known competing financial interests or personal relationships that could have appeared to influence the work reported in this paper.\par 
\vspace{12pt}
\hspace{-18pt}\textbf{Acknowledgments}\par
\vspace{12pt}
RB acknowledges IIT Tirupati for support through the new faculty seed grant (NFSG), and also the computational support.

\par
\vspace{12pt}
\hspace{-18pt}\textbf{Appendix A: Supplementary Material}\par
\vspace{12pt}
See supplementary information for the density of states, D($\omega$) and the spectral density of states, $\rho(\omega)$ of bulk water and methane-water with the central water molecule displacement $\ge$ 0.01 \AA. The supplementary information also contains D($\omega$) and $\rho(\omega)$ of both bulk water and only water active case of methane-water, and their differences, $\Delta D(\omega)$ and $\Delta \rho(\omega)$, measured as methane-water relative to bulk water. We present D($\omega$) and $\rho(\omega)$ of bulk water and methane-water systems, including the imaginary frequencies, and with different distance and angle cut-offs. Finally, We present the symmetric and anti-symmetric decomposition of modes for bulk water system.

\onecolumngrid
\clearpage
\appendix
\section{Supplementary Material}\label{sinfo}
\vspace{12pt}
In Figure \ref{figs1}, we represent the detailed analysis of the density of states, D($\omega$) and spectral density of states, $\rho(\omega)$ of bulk water and methane-water systems for all the normal modes having displacement of central water molecule $\ge$ 0.01 \AA. We also present D($\omega$) and $\rho(\omega)$ of both bulk water and only water active case of methane-water, and their differences, $\Delta D(\omega)$ and $\Delta \rho(\omega)$ in Figure \ref{figs2}. The $\Delta D(\omega)$ and $\Delta \rho(\omega)$ are measured by subtracting the bulk water response from the methane-water response.  In Figure \ref{figs3}, we present D($\omega$) and $\rho(\omega)$ of bulk water and methane-water systems, including the nearly 7\% imaginary frequencies for all the clusters. Figure \ref{figs4} contains the spectral density of states, $\rho(\omega)$ for clusters with central water molecule \mbox{O-H} displacement cut-offs ranging between 0.01 \AA\ - 0.05 \AA\ and central water molecule HOH angle displacement cut-offs between 1$^{\circ}$ - 5$^{\circ}$. In Figure \ref{figs5}, we decompose the symmetric and anti-symmetric modes based on the central water molecule \mbox{O-H} displacement for bulk water clusters.
\renewcommand{\thefigure}{S1}
\begin{figure}[htb!]
\centering
\includegraphics[width=0.7\linewidth]{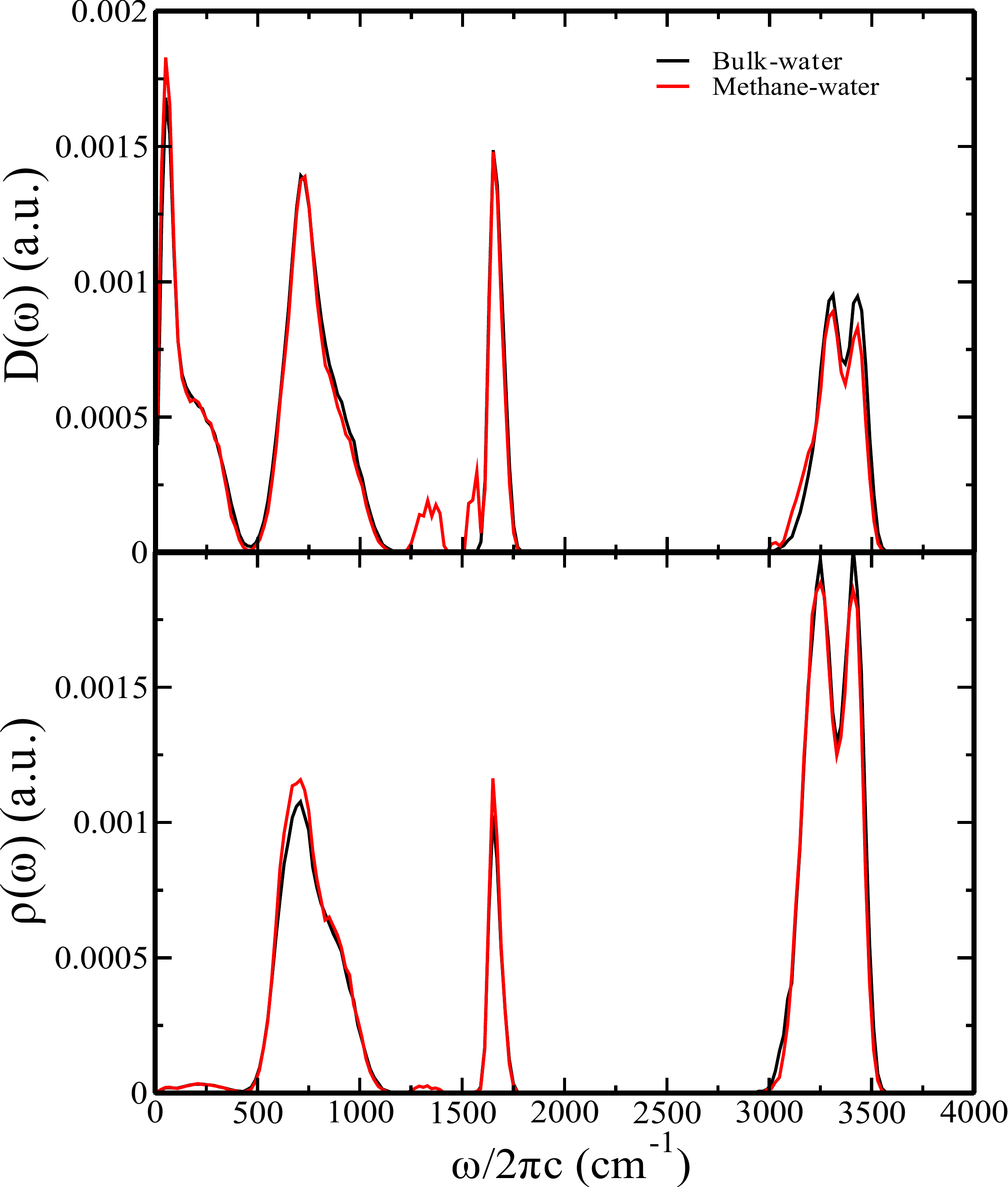}
\caption{\label{figs1}The density of states, D($\omega$) and the spectral density of states, $\rho(\omega)$ of bulk water (black) and methane-water (red) systems with central water molecule \mbox{O-H} displacement $\ge$ 0.01 \AA}
\end{figure}

\renewcommand{\thefigure}{S2}
\begin{figure}[htb!]
\centering
\includegraphics[width=0.7\linewidth]{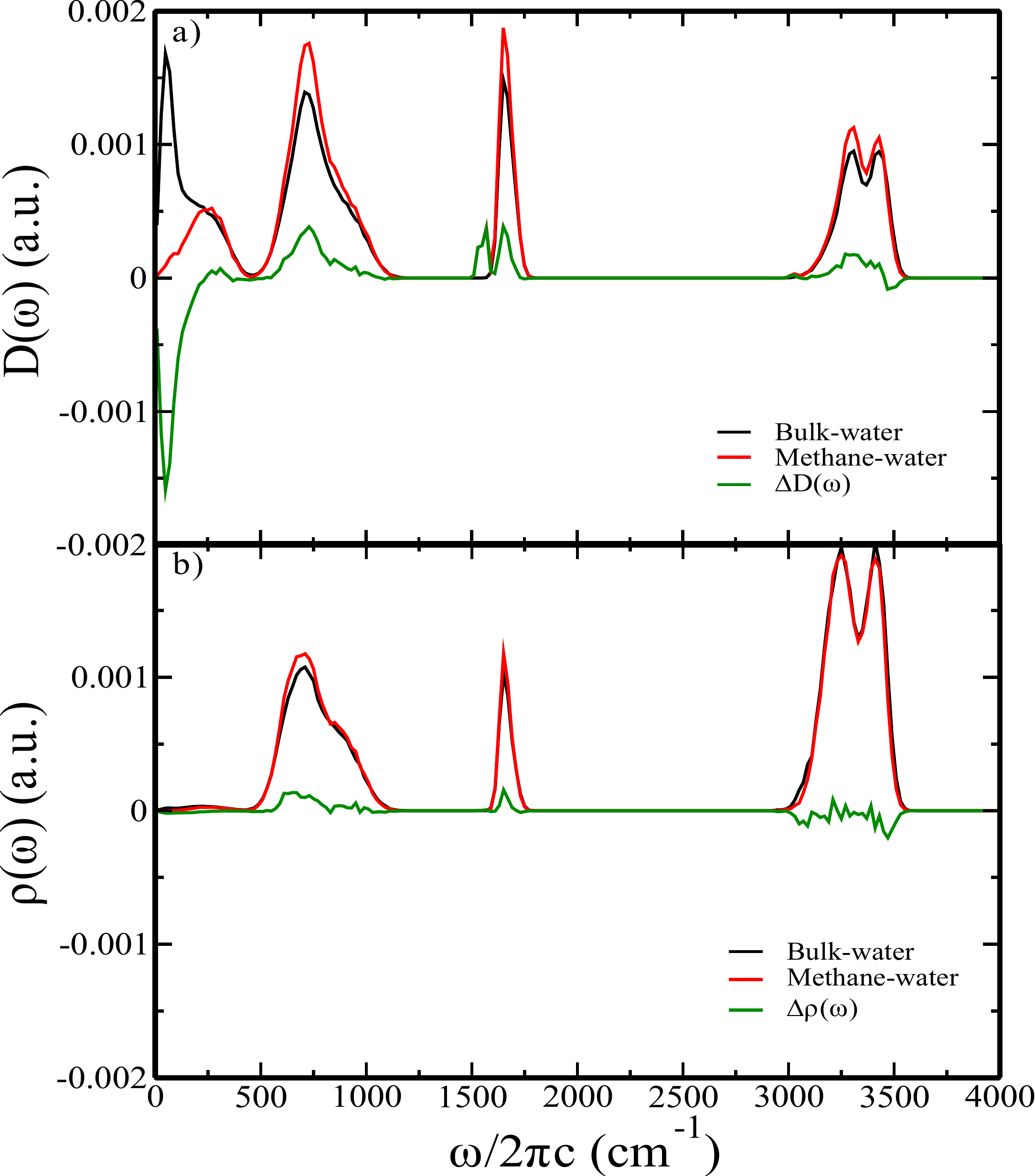}
\caption{\label{figs2} The density of states, D($\omega$) of bulk water (black), only water active modes of methane-water system (red) and the difference in density of states, $\Delta$ D($\omega$) of only water active modes of methane-water response relative to bulk water response (green). (b) The spectral density, $\rho(\omega)$ of bulk water (black), only water active modes of methane-water system (red) and the difference in spectral density, $\Delta\rho(\omega)$ of only water active response relative to the bulk water response (green).}
\end{figure}

\renewcommand{\thefigure}{S3}
\begin{figure}[htb!]
\centering
\includegraphics[width=0.7\linewidth]{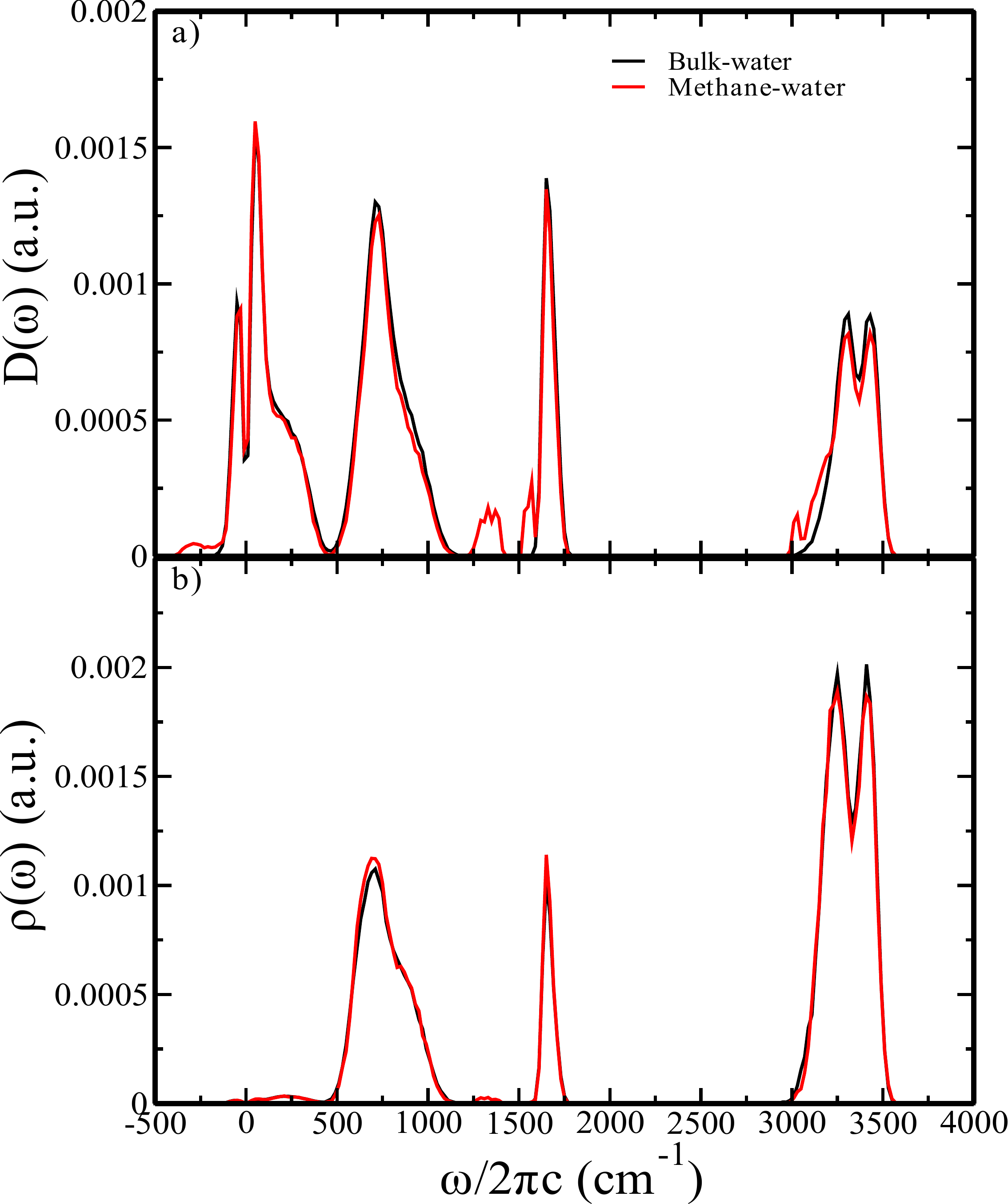}
\caption{\label{figs3} The density of states, D($\omega$) and spectral density of states, $\rho(\omega)$ of bulk water (black) and methane-water (red) with imaginary frequencies. }
\end{figure}

\renewcommand{\thefigure}{S4}
\begin{figure}[htb!]
\centering
\includegraphics[width=0.7\linewidth]{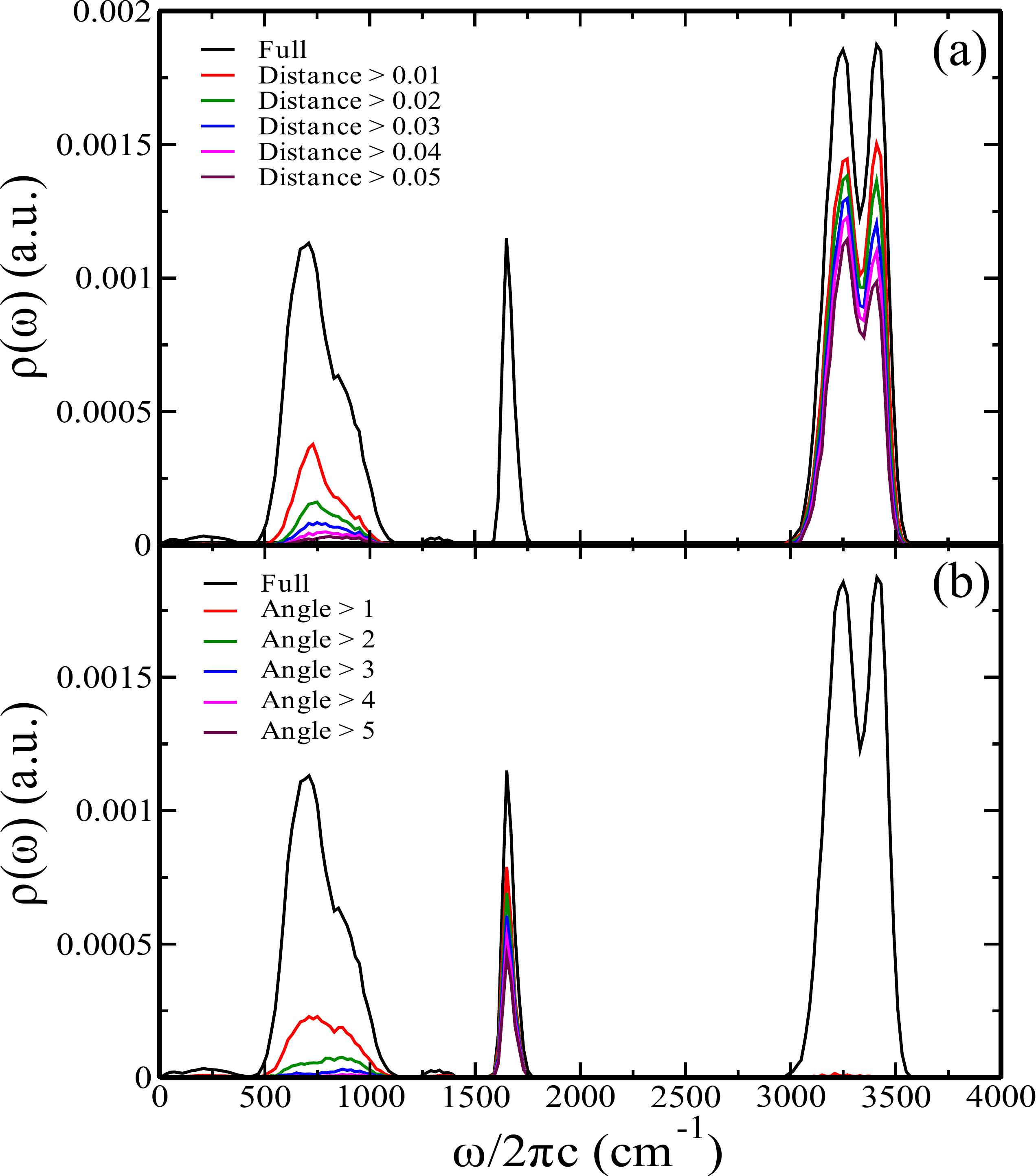}
\caption{\label{figs4} The density of states, D($\omega$) of  methane-water clusters with different distance cut off (in \AA) and angle cut off (in degrees).}
\end{figure}

\renewcommand{\thefigure}{S5}
\begin{figure}[htb!]
\centering
\includegraphics[width=0.7\linewidth]{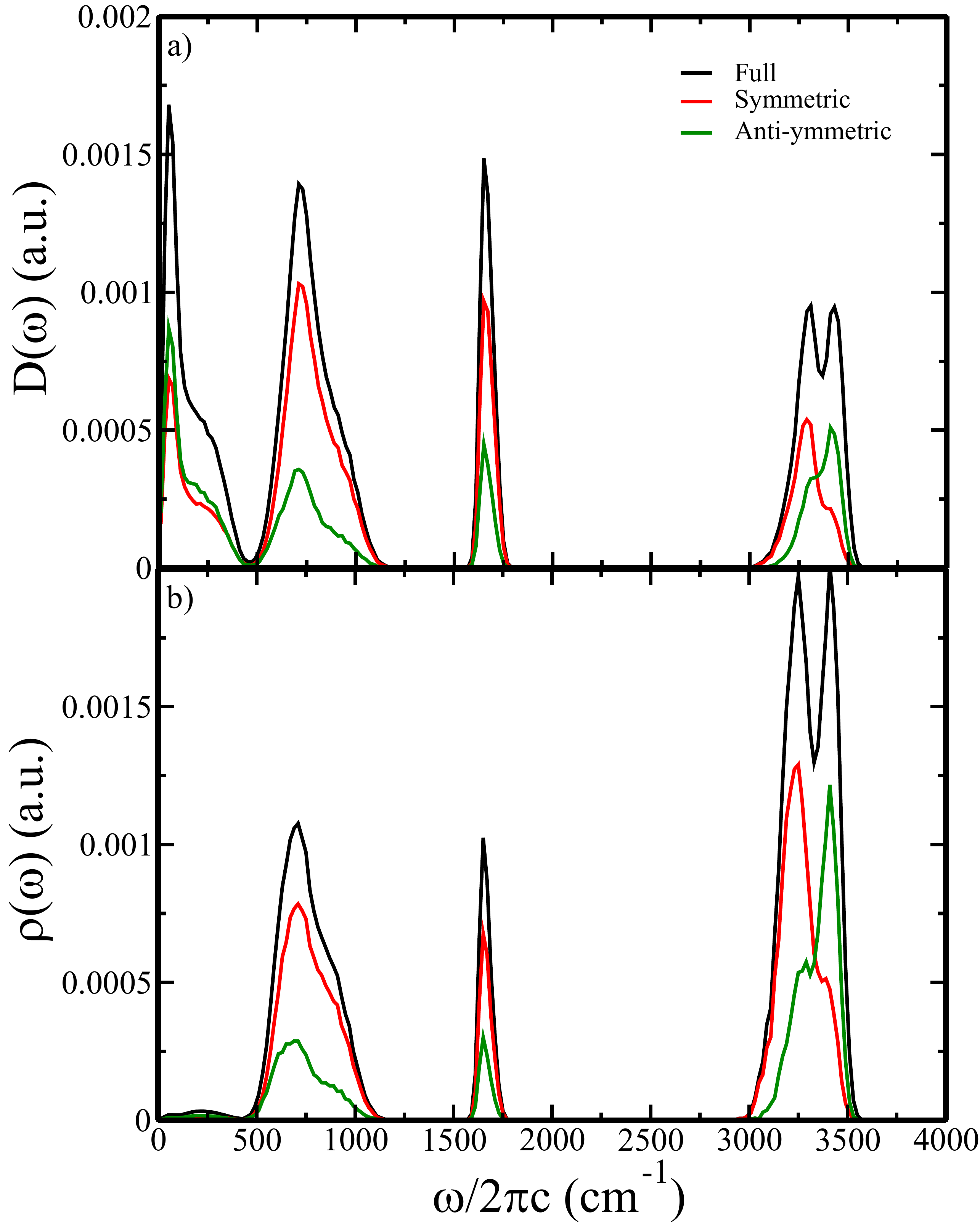}
\caption{\label{figs5} The density of states, D($\omega$) and the spectral density of states, $\rho(\omega)$ of bulk water clusters (black), with symmetric (red) and anti-symmetric (green) modes decomposed .}
\end{figure}
\clearpage
\end{document}